\documentclass[epsf,useAMS,usenatbib,usegraphicx]{mn2e}
 
\usepackage{amssymb,amsmath}
\usepackage{epsfig}
\usepackage{rotating}

\title[Investigating stellar-mass black hole kicks]
  {Investigating stellar-mass black hole kicks}
\author[S.~Repetto, M.B.~Davies and S.~Sigurdsson]
  {Serena~Repetto$^{1,2}$\thanks{E-mail:
S.Repetto@astro.ru.nl},~~Melvyn~B.~Davies$^1$~~and~~Steinn~Sigurdsson$^3$\\
  $^1$Lund Observatory, Department of Astronomy and Theoretical Physics, 
  Box 43, SE$-$221 00 Lund, Sweden\\
    $^2$Department of Astrophysics/IMAPP, Radboud University Nijmegen, P.O. Box 9010, 6500 GL Nijmegen, The Netherlands\\
$^3$Department of Astronomy \& Astrophysics, Pennsylvania State University, 525 Davey Lab, University Park, PA 16802, USA
}

\date{\today}

\pagerange{\pageref{firstpage}--\pageref{lastpage}} \pubyear{0000}

\def\LaTeX{L\kern-.36em\raise.3ex\hbox{a}\kern-.15em
    T\kern-.1667em\lower.7ex\hbox{E}\kern-.125emX}

\usepackage[colorlinks=false]{hyperref}

\begin{document}

\label{firstpage}

\maketitle

\begin{abstract}
We investigate whether stellar-mass black holes have to receive natal kicks in order to 
explain the observed distribution of low-mass X-ray binaries containing black holes within
our Galaxy.
Such binaries are the product of binary evolution, where the massive
primary has exploded forming a stellar-mass black hole, probably after 
a common envelope phase where the system contracted down to separations
of order $10-30~{\rm R}_\odot$.
We perform population synthesis calculations of these binaries, applying both kicks
due to supernova mass-loss and natal kicks to the newly-formed
black hole. We then
integrate the trajectories of the binary systems within the 
Galactic potential. We find that natal kicks are in fact necessary
to reach the large distances above the Galactic plane achieved by some binaries.
Further, we find that the distribution of natal kicks would seem to be similar to that of neutron
stars, rather than one where the kick velocities are reduced by the ratio
of black hole to neutron-star mass (i.e. where the kicks have the same momentum).
This result is somewhat surprising; in many pictures of stellar-mass black-hole 
formation, one might have expected black holes to receive kicks having the same 
momentum (rather than the same speed) as those given to neutron stars.
\end{abstract}


\section{Introduction}
It has long been known that neutron stars receive kicks at birth in 
the range $\sim 200-400$ km/s  (so called {\em {natal kicks}}),
when they are formed in core-collapse supernovae, for example via proper motion
studies of pulsars 
(\citealt{cord}, \citealt{lyn}).  
Whether stellar-mass black holes (for brevity, black holes hereafter) receive these kicks too 
is still a matter of debate. Black holes can be studied via interacting X-ray binaries
which contain them.  There are several known X-ray binaries which are known
to contain black holes or contain black hole
candidates (\citealt{jn}, \citealt{Oz}).
In these systems, the massive primary has evolved to form a black hole via a core-collapse
supernova and material is currently flowing from the lower-mass secondary
(typically via Roche-lobe overflow) onto the black hole via an accretion disc
(for a detailed review on the evolution of compact binaries see \citealt{tau}).
When the primary explodes as a supernova, the mass loss from the system
can unbind the binary or at least give it a kick (as the mass lost has a net momentum
in the rest frame of the binary). In addition, any natal kick received by the black 
hole will affect the orbital properties of the binary and its orbit within the Galaxy.
By studying the orbit of a binary, or even its location within the Galaxy, one might
obtain a limit on the range of allowed natal kicks. 
A number of studies have considered the 
motion of individual binaries within the Galaxy.

\citealt{steinn} considered GRO J1655--40 (Nova Sco) and concluded that a natal kick more easily accounted for the high space velocity of the binary.

\citealt{Nel} studied Cygnus X-1 and concluding that a natal kick was not necessary 
to explain its space velocity.

\citealt{wil} considered GRO J1655--40 and suggested that
although a natal kick is not formally required to produce the system as observed today,
the inclusion of a (modest) natal kick more readily explains the system. They also
placed an upper limit on the natal kick of $\simeq 210$ km/s.

\citealt{dha} considered GRS 1915+105. They concluded that any
peculiar motion of the binary was more likely due to later scattering within
the Galactic disc than a natal kick when the black hole formed.

The binary XTE J1118+480 is located at 
a very high latitude (1.5 kpc above the Galactic disc, see \citealt{rem}) and it has a high space velocity (\citealt{mir}). \citealt{gua}
 concluded that
for this system a black hole natal kick was {\em required}. More recently, 
\citealt{fra} placed the value of the natal kick in the range 80 - 310 km/s.

\citealt{won} considered Cygnus X-1. They found that in this 
case the black hole progenitor could have received a relatively small 
natal kick (few tens of km/s with an upper limit of  77 km/s). If the system 
originated in the Cyg OB3 association (\citealt{mr}) , then the upper 
limit on a kick is reduced to 24 km/s. 

In this Paper, we consider the population of black hole X-ray binaries as a whole
(following the approach of  \citealt{wp}, \citealt{jn}, \citealt{zuo})
rather than consider the kinematics of an individual system. 
We synthesize a population of black-hole low-mass X-ray binaries (BH-LMXBs), using various
natal kick distributions, and integrating the systems within the Galactic potential.
We then compare their locations within
the Galaxy to a catalogue of known black-hole X-ray binaries having 
measured distances (\citealt{Oz}).

The paper is arranged as follows. 
We review  the current state of observations of X-ray binaries containing
either neutron stars or black holes in Section \ref{sec:obs}.
Our treatment of the motion of stars in the Galactic potential is given in Section \ref{sec:dyn}. 
In  Section \ref{sec:bkt} we review the effects of both natal
and supernovae mass-loss kicks on binaries.
In Section \ref{sec:bps} 
we present the results of the binary population synthesis 
which we discuss in Section \ref{sec:discussion}. 
The paper is concluded in Section \ref{sec:concl}.

\section{The Observed Binaries}
\label{sec:obs}
In our Galaxy there are $16$ dynamically confirmed black holes in LMXBs
and $33$ NS-LMXBs, whose distance and Galactic position is known;
see respectively \citealt{Oz}, \citealt{jn}, and references therein
(in particular, Jonker \& Nelemans consider only NSs not found in globular clusters).
We present the binaries in Tables \ref{tab:tab1} and \ref{tab:tab2},
along with their angular distribution, their distance from the Sun and their position, 
both in Galactic coordinates and in cylindrical ones
($R$ refers to the radial distance from the Galactic centre and $z$ refers to the distance from the Galactic plane). 
Concerning the black hole binaries, uncertainty on the distance is taken from \"Ozel et al.; for the neutron star binaries,
Jonker \& Nelemans calculated
the distance assuming two different Eddington peak fluxes, getting a maximum and a minimum value for the distance, of which we 
take the median value. 
\setcounter{table}{0}
\begin{table*}
\begin{minipage}{115mm}
\caption{{Observed properties of BH-LMXBs.}}
\label{tab:tab1}
\begin{tabular}{l c c c c c c c}
\hline
Name& l & b & d & $\Delta d$ &  R& z & Ref.\\
& \small{(deg)} & \small{(deg)} & \small{(kpc)} & \small{(kpc)} & \small{(kpc)} & \small{(kpc)} & \small{(distance)}\\
\hline
4U1543-47&330.0&+5.4&7.5& 0.5 & 3.92&0.70&[1]\\
XTEJ1550-564&325.9&-1.8&4.4 &  0.5 &    5.0&-0.14&[2]\\
GROJ1655-40 & 345.0 & +2.5 & 3.2 &0.5     & 4.98 &0.13 & [3]\\
1659-487 & 338.9 & -4.3 & 9.0  & 3.0      & 3.25 & -0.67 &[4]\\
1819.3-2525 & 6.8 & -4.8 & 9.9   & 2.4  & 2.14 & -0.82 &[5]\\
GRS1915+105&45.4&-0.2&9.0  & 3.0  &6.62&-0.03&[6]\\
GS2023+338&73.1&-2.1&2.39    & 0.14  &7.65&-0.09&[7]\\
GROJ0422+32&166.0&-12.0&2.0    & 1.0    &9.91&-0.41&[8]\\
A0620-003&210.0&-6.5&1.06     & 0.12   &8.92&-0.12&[9]\\
GRS1009-45&275.9&+9.4&3.82    & 0.27       &8.48&0.62&[10]\\
XTEJ1118+480&157.6&+62.3&1.7     & 0.1    &8.73&1.50&[11]\\
1124-683&295.3&-7.1&5.89     & 0.26       &7.63&-0.73&[10]\\
XTEJ1650-500&336.7&-3.4&2.6     & 0.7     &5.71&-0.15&[12]\\
1705-250&358.2&+9.1&8.6   & 2.1     &0.55&1.36&[13]\\
XTEJ1859+226&54.1&+8.6&8.0  & 3.0  &7.23&1.20&[10]\\
GS2000+251 & 63.4 & -3.0 & 2.7 & 0.7           &7.21 & -0.14 & [13]\\
\hline \\
\end{tabular}
\newline

References (from \citealt{Oz}): [1] Orosz 2010 private communication, [2] \citealt{oronew}, [3] \citealt{hje}, [4] \citealt{hyn}, [5] \citealt{oro2}, [6] \citealt{fen}, [7] \citealt{mil},
[8] \citealt{web}, [9] \citealt{can}, [10] \citealt{hyn}, [11] \citealt{gel}, [12] \citealt{hom1999}, [13] \citealt{bar}.
\end{minipage}
\end{table*}

\begin{figure}
\begin{center}
\epsfxsize=\columnwidth\epsfbox{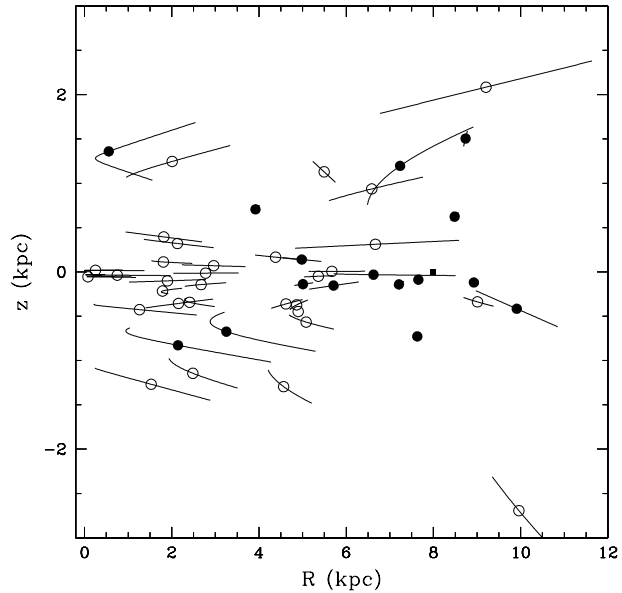}
\caption{Galactic distribution of NS-LMXBs (open circles) and BH-LMXBs (filled circles). 
The  radial distance from the Galactic centre
is $R = \sqrt{x^2 + y^2}$, where $x$ and $y$ are the cartesian coordinates in the 
Galactic plane, whereas the distance from the plane of the Galaxy is $z = d\sin{b}$.
Two neutron star binaries fall off the figure: 
XTEJ2123-058 and CygX-2. 
Solid lines account for the uncertainty of the distance from the Sun for each binary. The Sun is indicated as a square.}
\label{fig:obssample}
\end{center}
\end{figure}

\setcounter{table}{1}
\begin{table*}
\begin{minipage}{115mm}
 \caption{Observed properties of NS-LMXBs.}
\label{tab:tab2}
\begin{tabular}{l c c c c c c c}
\hline
Name& l & b & d & $\Delta d$ &  R& z & Ref.\\
& \small{(deg)} & \small{(deg)} & \small{(kpc)} & \small{(kpc)} & \small{(kpc)} & \small{(kpc)} & {\small{(distance)}} \\
\hline
EXO0748-676 & 279.98 & -19.81 & 7.95 & 2.3 & 9.96&-2.70&[1]\\
2S0918-54&275.85&-3.84& 5.05 & 1.5 & 9.0&-0.34&[2]\\
Cir X-1&322.12&0.04&9.15 &2.7 &5.67&0.00&[3]\\
4U1608-522&330.93&-0.85& 3.3 & 1.0 &5.36&-0.04&[4]\\
Sco X-1&350.09&23.78& 2.8 & 0.3 & 5.49& 1.13&[5]\\
4U1636-53&332.91&-4.82& 4.3 & 1.2 & 4.62&-0.36&[6]\\
4U1658-298&353.83&7.27& 9.85 & 2.9 & 2.01&1.25&[7]\\
4U1702-429&343.89&-1.32& 6.2 & 1.8 & 2.67&-0.14&[8]\\
4U1705-44&343.32&-2.34&8.4 & 2.4 & 2.40&-0.34&[8]\\
XTEJ1710-281&356.36&6.92&17.3 & 5.0 & 9.20&2.08&[8]\\
SAXJ1712.6-3739&348.93&0.93&6.9 & 2.0 & 1.81&0.11&[9]\\
1H1715-321&354.13&3.06&6.0 &1.8 & 2.13&0.32&[10]\\
RXJ1718.4-4029&347.28&-1.65&7.5 &2.2 & 1.79&-0.21&[11]\\
4U1728-34&354.30&-0.15&5.3 &1.6 & 2.78&-0.01&[12]\\
KS1731-260&1.07&3.65&6.2 &1.8 & 1.81&0.39&[13]\\
4U1735-44&346.05&-6.99&9.4 &2.8 & 2.48&-1.14&[8]\\
GRS1741.9-2853&359.96&0.13&7.75 & 2.3 & 0.25&0.01&[14]\\
2E1742.9-2929&359.56&-0.39&8.05 &2.3 & 0.08&-0.05&[8]\\
SAXJ1747.0-2853&0.21&-0.24&8.75 &2.5 & 0.75&-0.04&[15]\\
GX3+1&2.29&0.79&5.05 &1.5 & 2.96&0.07&[16]\\
SAXJ1750.8-2900&0.45&-0.95&6.1 &1.8 & 1.90&-0.10&[17]\\
SAXJ1752.3-3138&358.44&-2.64&9.25 &2.7 & 1.26&-0.43&[18]\\
SAXJ1808.4-3658&355.38&-8.15&3.15 &0.9 & 4.90&-0.45&[19]\\
SAXJ1810.8-2609&5.20&-3.43&5.95 &1.7 & 2.15&-0.36&[20]\\
4U1812-12&18.06&2.38&4.0 &1.2 & 4.38&0.17&[21]\\
XTEJ1814-338&358.75&-7.59&9.6 &2.8 & 1.53&-1.27&[22]\\
GX17+2&16.43&1.28&13.95 &4.1 & 6.67&0.31&[23]\\
SerX-1&36.12&4.84&11.1 &3.2 & 6.59&0.94&[8]\\
Aq1X-1&35.72&-4.14&5.15 & 1.5 &4.86 &-0.37&[24]\\
4U1857+01&35.02&-3.71&8.75 &2.5 & 5.08&-0.57&[25]\\
4U1916-053&31.36&-8.46&8.8 &2.6 & 4.56&-1.29&[26]\\
XTEJ2123-058&46.48&-36.20&18.35 &5.3 & 10.96&-10.83&[27]\\
Cyg X-2&87.33&-11.32&13.35 &3.9 & 15.02&-2.62&[28]\\
\hline \\
\end{tabular}
\newline
References (from \citealt{jn}): [1] \citealt{got}, [2] \citealt{jon}, [3] \citealt{ten}, [4] \citealt{mur}, [5] \citealt{bra}, [6] \citealt{fuj}, [7] \citealt{wij},
[8] \citealt{gal}, [9] \citealt{coc3}, [10] \citealt{taw}, [11] \citealt{kap}, [12] \citealt{bas}, [13] \citealt{mun}, [14] \citealt{jn}, [15] \citealt{nat}, 
[16] \citealt{kul}, [17] \citealt{kaa}, [18] \citealt{coc4}, [19] \citealt{int}, [20] \citealt{nat}, [21] \citealt{coc2}, [22] \citealt{str}, 
[23] \citealt{kul}, [24] \citealt{jn}, [25] \citealt{che}, [26] \citealt{gal}, [27] \citealt{hom1999}, [28] \citealt{sma}.
\end{minipage}
\end{table*}
Using the values in Table \ref{tab:tab1} and Table \ref{tab:tab2}, we plot the Galactic distribution of the binaries in Figure \ref{fig:obssample}. 
In representing the observed system on the $(R,z)$ plane we propagate the uncertainty on the distance into an uncertainty on $R$ and on $z$, the corresponding range of values for $R$ and $z$ being represented as a solid line.
The $z$-distribution 
of BH-LMXBs appears similar to the NS-LMXBs one 
(as already pointed out by \citealt{jn}, who calculated the rms z for the two samples); the {\em{Kolmogorov-Smirnov}} probability $P$ for the two distributions 
to be the same is indeed convincing: $P \sim 0.81$ 
(that increases to $0.85$ if binaries located at $z>2.0$ kpc are excluded from the test). Concerning the $R$ distribution, we may observe that NS-LMXBs seem to be more concentrated towards the Galactic centre 
with respect to the BH-LMXBs. This is very likely an observational bias, as already suggested by \citealt{jn},
since the BH binaries that we consider are those for which a dynamical measurement of the BH mass exists,
and this biases the binaries towards  those closer to us. 

\section{Integrating orbits within the Galaxy}
\label{sec:dyn}
In this section, we will give 
a short overview of how the orbital trajectories within  the Galaxy are calculated. 
For the Galactic potential, we make use of  the model proposed
 by \citealt{pac}. 
The  potential is modeled as the superposition of  three components due to the
  disc  ($\Phi_{\rm d}$), the spheroid (${\Phi}_{\rm s}$) and the
  halo ($\Phi_{\rm h}$), as given below
 \begin{equation}
{\Phi}_{\rm d}(R,z)=-\frac{G{\it M}_{\rm d}}{\sqrt{R^2+\left ({\it a}_{\rm d}+\sqrt{z^2+{\it b}_{\rm d}^2}\right )^2}}
\end{equation}
where $a_{\rm d}= 3.7$ kpc, $b_{\rm d}= 0.20$ kpc, and $M_{\rm d}=8.07 \times
10^{10} {\rm M}_\odot$.

\begin{equation}
{\Phi}_{\rm s}(R,z)=-\frac{G{\it M}_{\rm s}}{\sqrt{R^2+\left ({\it a}_{\rm s}+\sqrt{z^2+{\it b}_{\rm s}^2}\right )^2}}
\label{eq:spher}
\end{equation}
where $a_{\rm s}= 0.0$ kpc, $b_{\rm s}= 0.277$ kpc, and $M_{\rm s}=1.12 \times
10^{10} {\rm M}_\odot$.

\begin{equation}
\Phi_{\rm h}(r) = \frac{GM_{\rm c}}{r_{\rm c}}\left[\frac{1}{2}\ln\left({1+\frac{r^2}{r_{\rm c}^2}}\right)+\frac{r_{\rm c}}{r}\arctan\left({\frac{r}{r_{\rm c}}}\right)\right]
\end{equation}
where $r_{\rm c}= 6.0$ kpc and $M_{\rm c}=5.0 \times
10^{10} {\rm M}_\odot$. 

When integrating the trajectory of  a binary within the Galaxy, we make use
of the cylindrical symmetry of the potential. The equations of motion which 
are thus integrated are given below

\begin{equation}
\frac{dR}{dt}=v_R, \hspace{0.8cm}\frac{dv_R}{dt}=-{\left(\partial \Phi \over \partial R\right)_z}+\frac{j_z^2}{R^3}
\end{equation}
\begin{equation}
\frac{dz}{dt}=v_z, \hspace{0.8cm}\frac{dv_z}{dt}=-{\left(\partial \Phi \over \partial z\right)_R}
\end{equation}
where $R$ and $z$  are the cylindrical coordinates of the binary,
$j_z$ is the $z$-component of 
the angular momentum of the binary,  and $\Phi = \Phi_{\rm d}+\Phi_{\rm s}+\Phi_{\rm h}$.
 
\begin{figure}
\begin{center}
\epsfxsize=\columnwidth\epsfbox{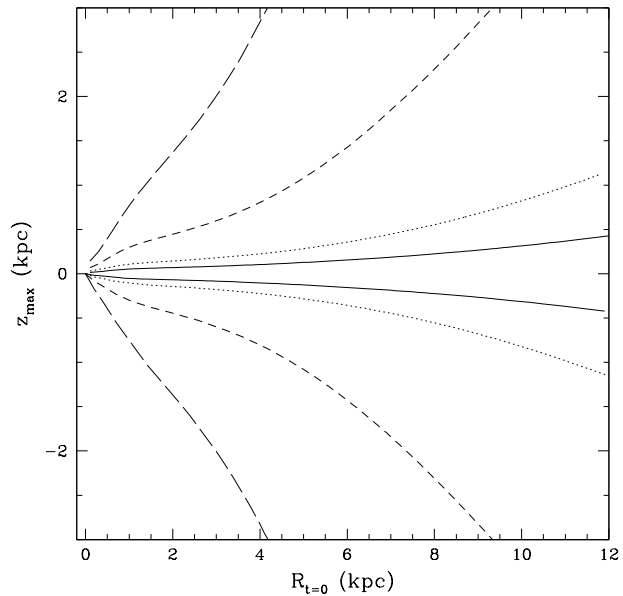}
\caption{Maximum $z$ reached by a binary over its trajectory. The object has been kicked perpendicularly to the Galactic plane, 
for four different magnitudes of the kick
($v_{\perp}=20$, $40$, $100$, $200$ km/s).}
\label{fig:zmax_with_20}
\end{center}
\end{figure}

It will turn out that typical kick velocities that the binary receives when the primary
explodes as a supernova are comparable to the circular orbital speed
in the Galaxy ($\sim 200$ km/s). This implies that kicks can significantly 
affect the trajectory of the binary within the Galaxy.
We can get an idea of the maximum $z$ reached by the binary as a
 function of the initial peculiar velocity.
We integrate the equations of motion of the binary for $\sim 10$ Gyrs (which is 
the typical main sequence MS-time of a sun-like star),
using a 4th-order Runge-Kutta integrator developed by SR,
and assuming the system was born right in the Galactic plane with a peculiar velocity perpendicular to the 
plane $v_{\perp}$. We perform 
the integration for different values of the velocity ($v_{\perp}=20$, $40$, $100$, $200$ km/s) 
and of the initial position $R_{t=0}$ over the plane, writing down 
the maximum $z$ reached over the trajectory (see Figure \ref{fig:zmax_with_20}). We see 
how $z_{max}$ is a rather strong function of
the initial position: for a fixed value of $v_{\perp}$, $z_{\max}$ gets smaller as 
the binary gets deeper in the potential well (i.e. smaller values
of $R_{t=0}$). It is clear from this figure that binary kick speeds in excess of 200 km/s will
be required in at least some of the observed systems shown in Figure 1.


\section{Kicks received by surviving binaries}
\label{sec:bkt}
In this section we consider the effects of the supernova explosion on the
binary.  We will see how the rapid mass loss from the supernova alone
could impart a kick on some systems whilst breaking others up. 
In addition, any kick imparted to the neutron star or black hole
on its formation (i.e. a natal kick) will also play a role, both in 
adding to the overall kick received by the binary, and in some
cases ensuring that the binary remains bound.\\
\indent
It is important to note that a {\em conspiracy of three velocities} will
 have an important role; namely the coincidence that 
the following three speeds are comparable:  
 the speed of
 a circular orbit in the Galaxy, the typical orbital speed within a 
 tight stellar binary when the primary explodes as a supernova, 
 and the characteristic kick speed the binary receives.
 This coincidence implies that kicks will significantly affect the orbit  of
 the binary within the Galaxy.

We begin by considering the case of zero natal kick. In other words, where
any kick is due solely to the rapid mass loss occurring during the supernova 
explosion. We will refer to this as the mass-loss kick $V_{\rm mlk}$ (also called {\em{Blaauw kick}}, \citealt{blaa}), which is given by
the expression below
\begin{equation}
V_{\rm mlk} = \frac{\Delta M}{M^\prime} \frac{M_2}{M} \sqrt{{\frac{GM}{a}}}
\label{eq:massloss}
\end{equation}
where $M$ is the total mass of the binary at the point of the supernova explosion,
$M^\prime$ is the total mass of the binary after the supernova explosion,
$\Delta M$ is the mass lost during the supernova explosion (i.e. $\Delta M = M -
M^\prime$), $M_2$ is the mass of the secondary, and $a$ is the binary semi-major
axis at the moment of the supernova explosion.

$V_{\rm mlk}$ can't be too large, 
since the binary must remain bound after the supernova: the mass loss 
must be less than half of the initial mass.
If we agree on a common envelope phase having shrunk the binary down to an orbital separation of $\sim10~{\rm R}_\odot$,
the resulting typical mass loss kicks for BH-LMXBs are of the order of $20-40$ km/s (for NS-LMXBs they are typically higher). 
Looking at figure \ref{fig:zmax_with_20}, we immediately realize how kicks of this size 
 cannot make the highest-z black hole binaries: 
in the optimal case of  $V_{\rm mlk} \simeq 40$ km/s 
 perpendicular to the Galactic plane,
 the maximum $z$ reached over the trajectory never exceeds $1$ kpc 
 (however, we do see binaries
 in the halo of our Galaxy at larger values of $z$, see table \ref{tab:tab1}).

We consider now the case where the neutron star or black hole produced in
the supernova receives a natal kick. 
If we assume the orientation of the natal kick is random with respect to the orbital plane, 
the natal kick $V_{\rm nk}$ combines with the mass loss kick $V_{\rm mlk}$ as given below:
\begin{equation}
V_{\rm k} = \sqrt{\left(\frac{M_{\rm bh}}{M^\prime}\right)^2 V_{\rm nk}^2 + V_{\rm mlk}^2 -2\frac{M_{\rm bh}}{M^\prime}V_{\rm nk,x}V_{\rm mlk}}
\end{equation}
where we have chosen the $x$ axis aligned with the orbital speed of the BH 
progenitor and the $y$ axis along the line connecting the two stars at the moment
of the supernova explosion.

Many distributions to model neutron star natal kicks have been proposed. 
For example, \citealt{hp} modeled the natal kick as a Maxwellian distribution
peaked at $300$ km/s. To solve the retention problem in globular clusters as well as the low eccentricity of
a subclass of Be X-ray binaries, 
two-peak distributions have also been proposed where one peak occurs at
a somewhat lower velocity (\citealt{bim1}, \citealt{bim2}).

\begin{figure}
 \begin{center}
\epsfxsize=\columnwidth\epsfbox{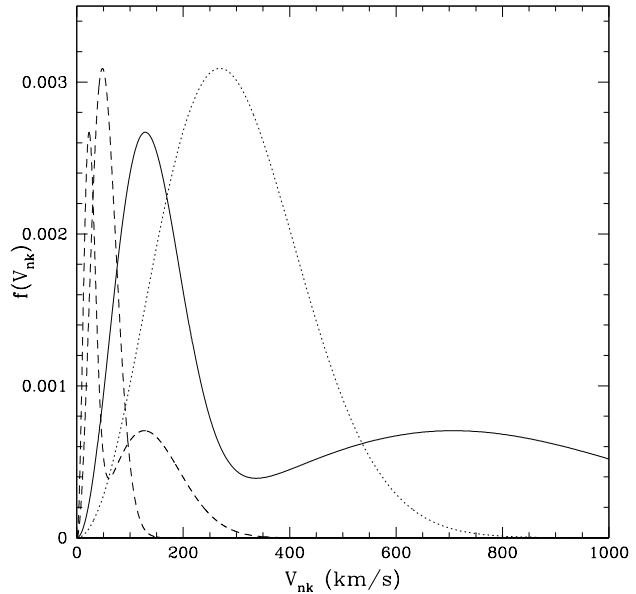}
 \caption{Natal Kick distributions used in our binary population
 synthesis calculations. Solid line corresponds to Arzoumanian distribution, dotted line to Hansen \& Phinney 
and the two dashed lines to these two distributions but with kick speeds
reduced, assuming that the momentum imparted to the black hole is the 
same as the momentum imparted to the neutron star.}
 \label{fig:NK1}
 \end{center}
\end{figure}

We consider two different natal kick distributions here: one is the 
 Hansen \& Phinney distribution, the other is a bimodal distribution
 proposed by  \citealt{arz} which has a lower peak at $\sim 100$ km/s and the 
higher peak at $\sim 700$ km/s.
We also consider modified versions of the above two distributions, 
which we term momentum-conserving kicks MCKs, where we assume that the momentum
imparted on a black hole is the same as the momentum given to a neutron star
using the two distributions. Thus the kick velocities will be reduced:
$V_{\rm nk,bh}=(M_{\rm ns}/M_{\rm bh})V_{\rm nk,ns}$.
For example, a $7~{\rm M}_{\odot}$ black hole receives a natal kick
 reduced by a factor of $5$:  for a neutron-star natal kick of $300$ km/s, the 
 black hole would receive a smaller kick of only  $60$ km/s. We show in figure
 \ref{fig:NK1} the natal kick distributions which we use.

It is important to recall that a  large fraction of binaries are broken up when
the primary explodes as a supernova.

Considering a population of binaries
where $M_1=11~{\rm M}_{\odot}$, $M_2=1.5~{\rm M}_{\odot}$, 
$M_{bh}=7.8~{\rm M}_{\odot}$, $a=10~{\rm R}_{\odot}$ ($M_1$ is the mass of the progenitor of the black hole, $M_2$ is the companion star and
$a$ is the pre-SN orbital separation), we impart the BH of each binary
a NK drawn randomly
from each of our four distributions. The fraction of 
systems remaining bound for each of the kick distributions is shown in Table  \ref{tab:surv}.
We also include the case where a $1.4~{\rm M}_\odot$ neutron star  is produced instead
of a black hole (taking as the progenitor mass $3.5~{\rm M}_{\odot}$).
A larger fraction of binaries remain bound for binaries containing black 
holes rather than neutron stars owing to the greater binding mass.
\setcounter{table}{2}
\begin{table}
\caption{Fraction of systems that stay bound after the SN.}
\label{tab:surv}
\begin{center}
\begin{tabular}{l l l}\hline
\multicolumn{3}{c}{Fraction of bound systems}\\
\hline 
& BH & NS \\
Hansen & $58\%$ & $30\%$\\
&  $56\%$$^a$ & $35\%$$^a$ \\
&  $64\%$$^b$ & $3\%$$^b$ \\
Bimodal & $54\%$ & $29\%$ \\
&  $56\%$$^a$ & $30\%$$^a$\\
&  $52\%$$^b$ & $10\%$$^b$\\
Hansen MCK & $99\%$ & -\\
Bimodal MCK & $95\%$ & -\\

\hline
\end{tabular}
\end{center}
$^{a}$ For a NK lying in the orbital plane.\\
$^{b}$ For a NK perpendicular to the orbital plane.\\
\end{table}

\begin{figure}
 \begin{center}
\epsfxsize=\columnwidth\epsfbox{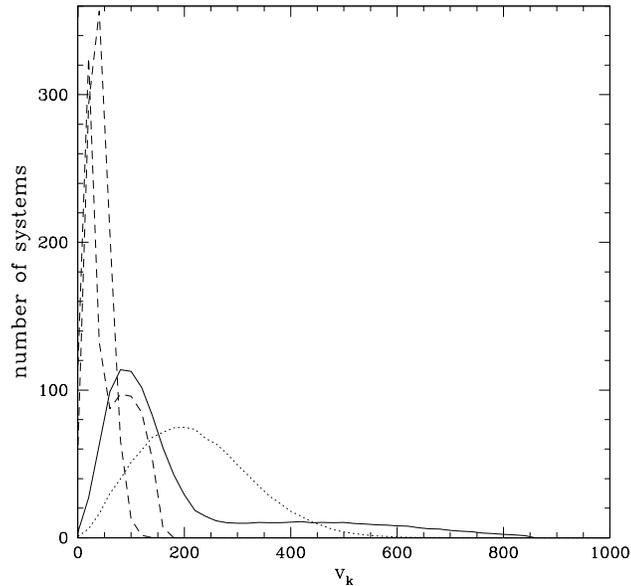}
 \caption{Peculiar velocity gained by the binary after an asymmetric supernova. Natal kicks have been drawn from Arzoumanian distribution
(solid line), 
Hansen \& Phinney distribution (dotted line), whereas the dashed lines correspond to the MCKs. The total number of systems for each curve has been normalized to 1000 and only systems that stay bound after the SN are represented.}
  \label{fig:typ}
 \end{center}
\end{figure}

We show in figure \ref{fig:typ} the distribution of  kick velocities for BH-LMXBs 
that we obtain drawing from each of the four natal kick distributions.  One should in particular note how 
the  kick velocities for the momentum conserving kicks are typically 
lower than $\sim 100$ km/s.

\section{Binary Population Synthesis}
\label{sec:bps}
In this section we discuss the calculation of the synthetic population
of black hole low-mass X-ray binaries. We produce a population
of BH-LMXBs considering their formation within the Galactic disc at a range of radii.
For each binary, we randomly draw the black hole natal kick, considering five different
natal kick distributions (including a zero natal kick) and give this kick a random direction
which we then add to the mass-loss kick due to the supernova to produce the total
kick speed of the binary $V_{\rm k}$. The gained velocity will be added randomly to the
circular velocity of the binary within the Galaxy. Each system is then integrated
forward in time within the Galactic potential for $\sim 3\times 10^{9}$ yr (which
is the MS-time of the $1.5~{\rm M}_{\odot}$ companion), and its position is
noted at random times over the trajectory. We are thus able to produce an entire population of 
BH-LMXBs given the initial distribution of progenitor systems in the Galactic
disc, their binary properties (separation and stellar masses) and the natal kick
distributions for the black holes formed.

We populate the disc of the Galaxy assuming the disc distribution of binaries to be 
proportional to the surface density of stars $\Sigma (R) \sim \Sigma_0 e^{-R/R_{\rm d}}$,
with a maximum distance from the Galactic centre of $R_{max}=10$ kpc.
We chose $R_{\rm d}$ to be the length scale of the thin disc of the Galaxy,  where
the progenitor systems are thought to be produced, $R_{\rm d} \sim 2.6$ kpc (\citealt{mcm}). 
Concerning the z-distribution of the binaries, we model it as an exponential
with scale height $\sim 0.167$ kpc (\citealt{bin}).

The population is formed by $100$ binaries with the following parameters: $M_{1}=
11~ {\rm M}_{\odot}$, $M_2=1.5~{\rm M}_{\odot}$, 
$M_{\rm bh}=7.8~{\rm M}_{\odot}$, $a=10~{\rm R}_{\odot}$ ($M_1$ is the mass of 
the progenitor of the black hole, $M_2$ is the companion star and
$a$ is the pre-SN orbital separation). For the black hole mass, we choose the average
 mass of stellar black holes in the Galaxy 
(see \citealt{Oz}); for the initial orbital separation, our choice is guided by
the typical results of common envelope evolution considered in detailed
binary evolution calculations for progenitor systems. 
We choose a typical mass loss in the supernova explosion of
 $\sim 3~{\rm M}_{\odot}$ (see \citealt{fry}), which delivers an associated 
mass-loss kick of $\sim 20$ km/s.

In addition, the black hole receives a natal kick drawn from one of the following five
distributions: 1) a natal kick of zero km/s; 2) one drawn from the Hansen \& Phinney 
distribution (\citealt{hp}); 3) one drawn from the bimodal distribution of \citealt{arz}; 4) as 2) but with the kick speed
multiplied by the factor $M_{\rm ns}/ M_{\rm bh}$; and 5) as 3) but 
with the kick speed multiplied by the factor $M_{\rm ns}/ M_{\rm bh}$.

\begin{figure}
\begin{center}
\epsfxsize=\columnwidth\epsfbox{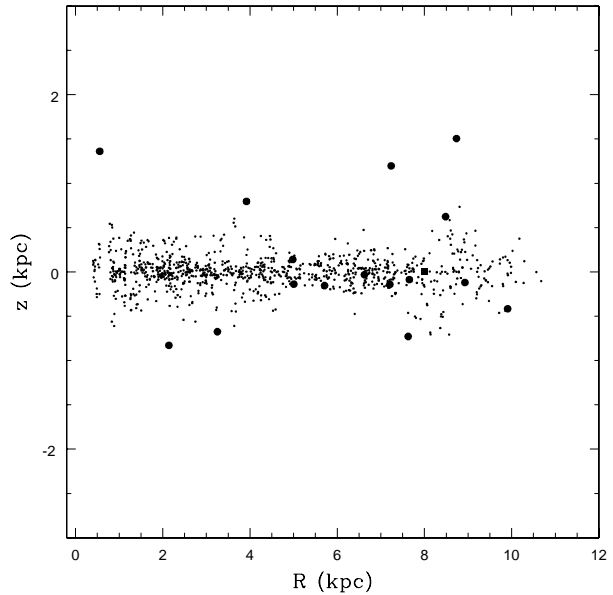}
\caption{Binary population synthesis for a sample of BH-LMXBs. No natal kick has been imparted to the BH. Smaller dots correspond
to the synthetic population, bigger ones to the observed binaries and the position of the Sun is denoted with a square.}
\label{fig:zero}
\end{center}
\end{figure}

\begin{figure*}
\begin{center}
\epsfxsize=1.8\columnwidth\epsfbox{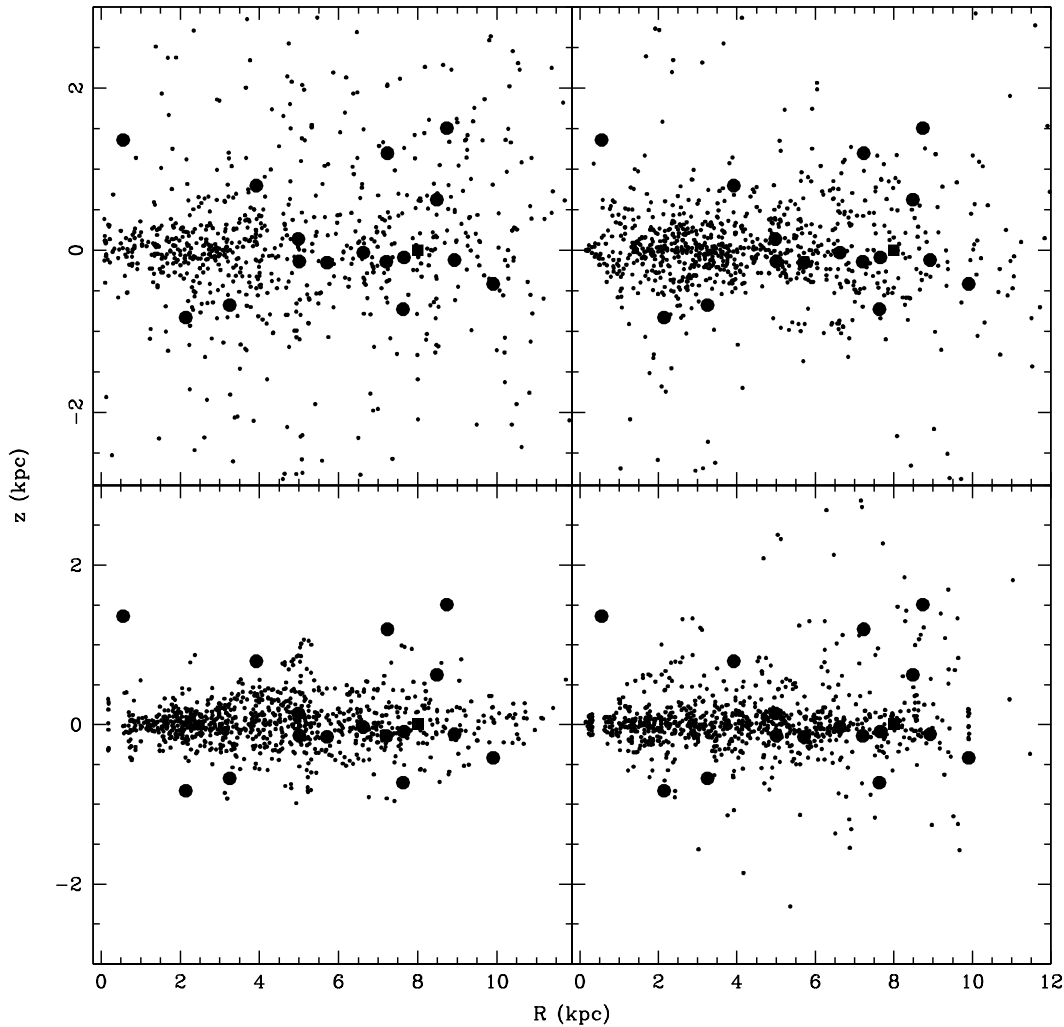}
\caption{Binary population synthesis for a sample of BH-LMXBs. Natal kicks have been drawn from Hansen \& Phinney (top left) distribution, a bimodal distribution (top right),
whereas the bottom figures correspond to the reduced natal kicks. Smaller dots correspond
to the synthetic population, bigger ones to the observed binaries and the position of the Sun is denoted with a square.}
\label{fig:adjoining}
\end{center}
\end{figure*}

We plot the positions of the 100 binaries -at random times of the trajectory- in Galactic 
cylindrical coordinates for zero black-hole natal kicks in Figure \ref{fig:zero}
and for the other four natal kick distributions in Figure \ref{fig:adjoining}.
From Figure \ref{fig:zero} it is clear that it is impossible to place BH-LMXBs seen at larger
values of $z$ when the black holes receive zero natal kicks. Either the Hansen \& Phinney
or the Arzoumanian distributions appear to fit the observed distribution of
BH-LMXBs, whereas those produced by natal kick distributions with velocities
reduced by a factor of $M_{\rm bh}/M_{\rm ns}$ (bottom panels in Figure \ref{fig:adjoining}) 
appear to produce distributions which are more concentrated on lower values
of $z$.

\subsection{Statistics of the results}
\begin{figure}
\begin{center}
\epsfxsize=\columnwidth\epsfbox{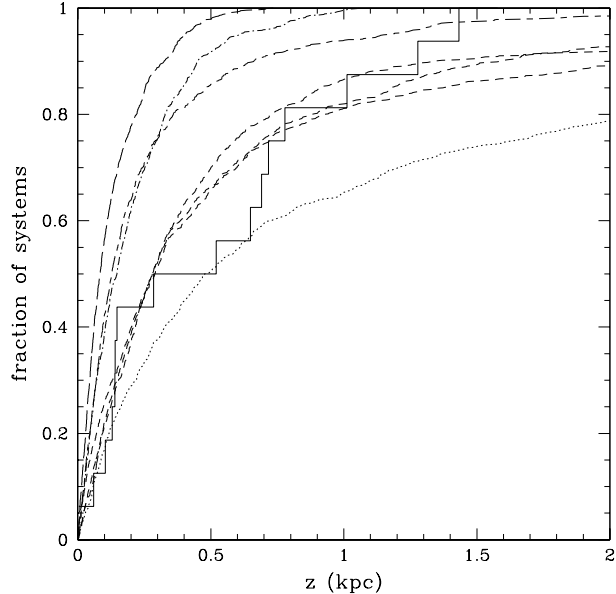}
\caption{Cumulatives which show the fraction of BH-LMXBs versus the distance from the Galactic plane, for the four different natal kicks
(dotted line is for an Hansen \& Phinney NK, dotted-dashed line is for a reduced Hansen \& Phinney NK, short-long-dashed is for a reduced Arzoumanian NK, whereas a zero 
NK scenario corresponds to the long-dashed line). For the Arzoumanian NK, we tested two additional scenarios,
a NK lying in the orbital plane and a NK perpendicular to it (see the three short-dashed lines).
Cumulatives are to be compared with the observed one (solid line).
}
\label{fig:BHcum}
\end{center}
\end{figure}

In order to quantify the results of the BPS, we show in figure \ref{fig:BHcum} the fraction of binaries that at some time 
over the trajectory are located at a distance $z$ from the Galactic plane less than a certain value. 
We include in the plot the results of the BPS for which no NK has been imparted to the black hole. 

\begin{figure}
\begin{center}
\epsfxsize=\columnwidth\epsfbox{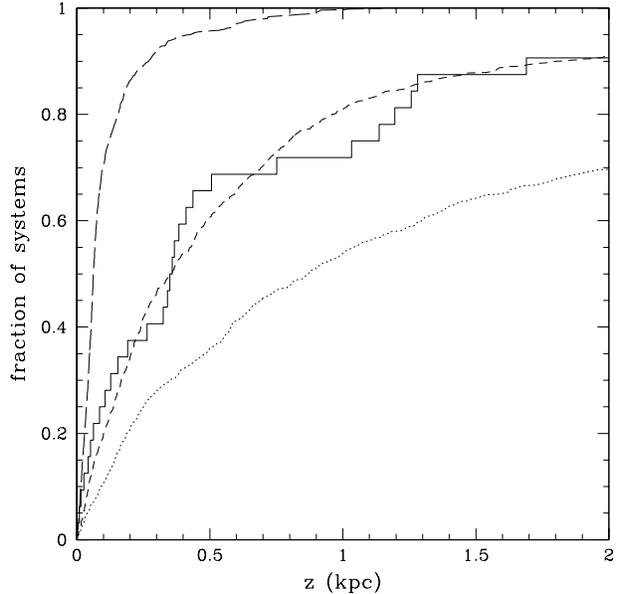}
\caption{Cumulatives which show the fraction of NS-LMXBs versus the distance from the Galactic plane, for the four different natal kicks
(dotted line is for an Hansen \& Phinney natal kick, short-dashed line is for an Arzoumanian natal kick, whereas a zero natal kick 
scenario corresponds to the long-dashed line). Cumulatives are to be compared with the observed one (solid line).
}
\label{fig:NScum}
\end{center}
\end{figure}

It is evident how the mass-loss kicks alone cannot account for the
 z-distribution of the observed binaries. 
A reduced Hansen \& Phinney NK cannot make the binaries that are located at $z\ga 1$ kpc;
in particular, the percentage of binaries that get to $z$ higher than $1$ kpc is only the $0.5\%$. 
With a reduced Arzoumanian NK the percentage gets higher ($\sim 6\%$), though the fit remains unsatisfactory (see table \ref{tab:prob}).
It then turns out to be very difficult, with a momentum conserving kick, to reproduce the binaries XTEJ118+480, 1705-250, XTEJ1859+226, 
which are located respectively at $z=1.5$, $1.36$, $1.20$ kpc. Section \ref{sec:discussion} is dedicated to the detailed study of these sources.\\
\indent
We are aware that the integration time we chose ($\sim 3\times 10^9$ years) might be higher than the actual age of some of the observed binaries,
particularly those whose mass transfer is driven by angular momentum losses, or those with a relatively massive companion star ($M_2\sim 3~{\rm M}_{\odot}$,
see the updated catalogue \citealt{rit}).
We then carry out other two synthesises for the BH-LMXBs, integrating their trajectories for $\sim 10^8$ years and for $\sim 5\times10^8$ years. In the first integration, 
the percentage of binaries that reach $z$ higher than $1$ kpc is $0\%$ for a reduced Hansen \& Phinney NK and $1\%$ for an Arzoumanian reduced NK. 
In the second integration, the percentages are respectively are $0.2\%$ and $2.2\%$. Concerning the KS-test,
the resulting probabilities get $1$ or $2$ order of magnitudes lower when choosing a reduced integration time;
this is easily explained, since the
binary doesn't live long enough to be seen at high $z$.\\
\indent
We wonder whether a larger mass-loss kick would affect our conclusions.
Referring to equation \ref{eq:massloss}, we see that the mass-loss kick increases
either in the case of a larger mass-loss, or a more compact initial binary, or a larger companion mass.
It is believed that in the SN event the Helium star loses no more than $3-4~{\rm M}_{\odot}$ 
(before exploding as a SN, the Helium star suffers from strong WR winds after the common envelope phase,
see for example \citealt{fry}). Concerning the initial orbital separation,
there is a limiting minimum value for which either one or both of the two stars fill their Roche lobe. 
The following parameters, 
$M_{1}=11~ {\rm M}_{\odot}$, $M_2=3.0~{\rm M}_{\odot}$, $M_{\rm bh}=7.8~{\rm M}_{\odot}$, $a=6~{\rm R}_{\odot}$,
give a recoil velocity $V_{\rm mlk}$ of $\sim 40$ km/s.
We then perform two synthesises in which we fix the mass loss kick to $\sim 40$ km/s,
testing the two types of reduced natal kicks. 
The corresponding KS probabilities remain unsatisfactory:
$2\times10^{-3}$  for a reduced Hansen NK, and $2\times10^{-2}$ for a reduced bimodal NK.
We shall also stress that the integration time for these two synthesises has been set to $\sim 3\times 10^9$ years; 
decreasing the integration time would make the KS probabilities even lower.\\
\indent
We perform a BPS for NS-LMXBs as well (binary parameters chosen: 
$M_1=3.5~{\rm M}_{\odot}$, $M_2=1.0~{\rm M}_{\odot}$, $M_{\rm ns}=
1.4~{\rm M}_{\odot}$, $a=7.0~{\rm R}_{\odot}$): 
in Figure \ref{fig:NScum} results are shown. It is pretty clear that
a bimodal distribution better fits the observed sample. This a strong case for neutron stars receiving
a bimodal NK at birth. Previous studies, focused on NSs in  Be X-ray binaries and double NS binaries
(see works by \citealt{bim2} and \citealt{doubleNS}), showed that at least some of the NSs should have received a lower kick at birth. We highlight our work 
as the first test of a bimodal distribution being a better fit to the Galactic position of NS low-mass X-ray binaries. 
When excluding from the test the observed NS binaries that are located at $z>2$ kpc, the KS probability rises to $0.19$; 
this is easily explained, since
the change in normalization shifts the simulated curve towards the observed one.\\
\indent
In table \ref{tab:prob} KS probabilities for the different types of scenario are shown
(the probabilities are reasonably accurate for our number of data points, see \citealt{press}).\\

\setcounter{table}{3}
\begin{table}
\caption{KS probabilities for the BPS.}
\label{tab:prob}
\begin{center}
\begin{tabular}{l c c}\hline
\multicolumn{3}{c}{KS probabilities}\\
\hline 
& BH-BPS & NS-BPS \\
\multicolumn{3}{c}{Integration time $3\times 10^9$ years}\\
Hansen NK & $0.20$ & $2.6\times10^{-3}$ \\
Bimodal NK & $0.18 $ & $0.78$ \\
Hansen MCK & $2\times10^{-3}$ & - \\
Bimodal MCK& $1\times10^{-2}$     & - \\
zero NK & $2\times 10^{-4}$ & $5\times10^9$ \\
\multicolumn{3}{c}{Integration time $5\times10^8$ years}\\
Hansen MCK &$7\times10^{-4}$ & -\\
Bimodal MCK &$5\times10^{-3}$ & -\\
\multicolumn{3}{c}{Integration time $10^8$ years}\\
Hansen MCK &$4\times10^{-4}$ & -\\
Bimodal MCK &$6\times10^{-4}$ & -\\
\hline
\end{tabular}
\end{center}
\end{table}

\section{Discussion}
\label{sec:discussion}

We now aim at deriving the minimum natal kick required to place the
observed BH-LMXBs in their current locations.  Rather than considering the properties
of a general progenitor binary system (i.e. stellar masses, black-hole mass,
and binary separation), we use the observed properties for each known system,
where possible, to more accurately calculate the mass-loss kick and the effect
of any natal kick on the particular system 
(see \citealt{rit} for an updated catalogue of LMXBs in the Galaxy).

For simplicity here, we consider a kick in the (optimal) direction
perpendicular to the Galactic disc. We first compute, via conservation of energy, the minimum kick
 $V_{\perp}$ for the binary 
to reach the current position ($R_0$, $z$) after traveling in the Galactic potential,
assuming the binary is born right over the Galactic plane at some 
radius $R_0$. Thus 
\begin{equation}
\frac{1}{2}{V_{\perp}}^2 + \Phi \left (R_0, 0 \right ) = \Phi \left ( R_0, z \right )
\end{equation}
where $\Phi$ is the gravitational potential of the Galaxy.

We show in Table \ref{tab:min} the minimum
perpendicular kick $V_\perp$ and minimum natal kick $V_{\rm nk}$
 required for those
BH-LMXB systems which have
 relatively well-constrained binary properties.
One should understand that the natal kick value quoted in Table 5 is the
absolute lower limit, assuming that the natal kick occurred in the perfectly optimal
direction. In practice, the natal kick required will be much larger in many realizations
(i.e. different kick directions for the natal and mass-loss kicks). For example,
 for many directions, the necessary natal kick to reproduce the current location
 of XTE J1118+480 is up to 300 km/s (as was also seen in \citealt{fra}).

The minimum peculiar velocity is normally greater than $80$ km/s for 
binaries that are located at $z>1$ kpc, 
or for binaries
that are closer in towards the Galactic center and at $z>0.6$ kpc. 
Particularly, for systems that are located at $R\la 3$ kpc from the Galactic center,
typical required velocities are greater than $100$ km/s for the highest-z systems. 
These high velocities cannot evidently be accounted for by a mass-loss kick alone.
In other words, the current location of at least some systems clearly requires
the presence of a black hole natal kick broadly in the range $100 - 500$ km/s.
For other systems, the current location could be reached with the black hole
having received no natal kick.  We note that our results are consistent with those
found earlier (\citealt{Nel},
\citealt{jn}, \citealt{wil}, \citealt{dha},
\citealt{fra}).

The system 1705-250 requires the largest minimum natal kick. This is because
the system is located close to the Galactic centre and therefore has climbed out
of a deeper potential well, assuming it was born at a comparable radius in the Galactic disc.
As there is a strong radial dependence of the Galactic potential this close in,
we compute the minimum peculiar velocity both launching  the binary out of the disc 
at $R=0.5$ kpc (the current location is at $R=0.55$ kpc, $z= 1.36$ kpc)
 and $R=2$ kpc. In the first case, 
we get $V_{\perp}\sim 400$ km/s, while in the second case a velocity 
$V_{\perp}\sim 250$ km/s is needed.
These velocities require a minimum natal kick of $440$ km/s 
and of $260$ km/s respectively.
Our results might have been affected by the choice of a spherically symmetric bulge
(i.e. $a_{\rm s}= 0.0$ kpc, see equation \ref{eq:spher} for the potential of the spheroid);
we then take a {\em{pseudobulge}} with $a_{\rm s}= 1.0$ kpc.
The resulting $V_{\perp}$ is $\sim320$ km/s for $R=0.5$ kpc and $V_{\perp}\sim 220$ km/s for $R=2.0$ kpc.
The associated minimum natal kicks are $340$ and $230$ km/s respectively.
For both of the two model for the Galactic potential, 
the required velocities are larger than the largest velocities drawn from a 
reduced-velocity kick, from either the Hansen \& Phinney or
Arzoumanian kick distributions.\\
\indent
We may wonder whether our conclusions on the required minimum natal kick
would be affected by a new estimation of the distance. 
For all of the previously mentioned four binaries, the distance
has been derived from the estimation of the absolute magnitude of the companion (see respectively \citealt{oro2002}, \citealt{oro2}, \citealt{gel}, \citealt{bar}).
\citealt{jn} observed that this method typically underestimates the distance. 
Also, we point out that any underestimation of the contribution of the disc to the observed magnitude of the companion star, 
would lead to an underestimation of the distance.
To quantify the effect of a new estimation of BH binaries distance,
we computed the minimum required natal kick, 
the distance being $10$, $25$, $50$, $100$ $\%$ larger than the nominal value. We perform the computation
for our four candidates of BHs receiving the same NK as NSs. 
The required minimum NK decreases
in all the cases except for XTEJ1118+480.
This is easily explained since a larger distance
would move the binary further out the Galactic potential
well. The binaries 4U1543-47, 1819.3-2525, 1705-250
require a minimum natal kick of $45$, $22$, $92$ km/s, respectively,
when the distance is increased of the $100\%$.
XTEJ1118+480 shows instead an increase of the minimum NK
up to $\sim 100$ km/s when the distance is multiplied by a factor of $2$.\\
\indent
An alternative scenario for the formation of BH binaries
would be via dynamical interactions in globular clusters (GCs).
However, it is still uncertain whether BHs
are retained in GCs or whether they follow
a different dynamical evolution than NS binaries. 
So far, no BH X-ray binary has been found in Galactic globular clusters
(\citealt{verbunt});
the strongest BH candidate in a GC is the one found
by \citealt{macca}, in the Galaxy NGC 4472.
The question whether black holes might be retained in a globular cluster
has been largely discussed in the literature
(e.g., \citealt{kulkarni}, \citealt{quist}, \citealt{porto}, \citealt{mh}).
It is generally thought that black holes would tend
to decouple dynamically from the rest of the cluster
and to segregate into the core,
where they would form BH-BH binaries.
Sequential dynamical interactions between these binaries 
and single BHs would lead to the ejection of the BHs from the cluster
in a timescale shorter than $\sim 10^9$ years. 
In case one black hole survives in the cluster,
it could potentially capture a stellar companion via two main mechanisms:
tidal capture of a star by the BH
or exchange interactions of the BH with a primordial binary (see \citealt{kalogera} and reference therein). 
Stars the BH is interacting with have a typical mass $\lesssim 1~M_{\odot}$ at this stage of the life of the GC.
It has been shown that the
two-body tidal capture scenario between a NS and a low-mass star
is likely to result in a merger
(\citealt{daviesGC}, \citealt{kumar}, \citealt{mcmillan}, \citealt{rasioGC}).
Specifically, \citealt{daviesGC}
showed that an encounter between a NS and a red-giant would result in a merger in some $70\%$ of the cases,
and some $50\%$ for an encounter between a NS and a MS-star.
For a BH, tidal forces are expected to be much larger,
so that a merger becomes even more likely.
Regarding the exchange interaction scenario,
we find it unlikely that the resulting BH-MS star binary
will get a large kick in subsequent dynamical encounters
for the binary to be expelled from the cluster.
Nevertheless, in the optimistic case that the binary managed to be ejected from the cluster
with a velocity comparable to the escape speed of the cluster,
we may compute the resulting overall distribution of Galactic BH binaries.
We assume the binaries to be born in a spheroid of $20$ kpc radius around the Galactic centre,
taking the halo distribution of \citealt{deh}. We then kick 
the binaries with a velocity of $45$ km/s and we follow their motion in the Galactic potential for $\sim 3\times 10^{9}$.
The resulting KS test gives probabilities lower than $8\times 10^{-7}$,
even when we double the BH binaries distance.
This result is not surprising
since we rarely get any binaries in the disc 
when assuming that all binaries are born in GCs.
It could be that this mechanism would work 
for the BH binaries found at the highest $z$. 
However, at least in the case of XTE J1118+480 
there are strong arguments for rejecting a GC origin.
\citealt{gua} estimated the age of the system,
using stellar evolution calculations,
to be between $2$ and $5$ Gyr,
rendering a globular cluster origin unlikely.
\citealt{her} performed a detailed chemical analysis of the optical star.
Starting from different
initial metallicities of the companion, they calculated 
the expected abundances
after contamination from SN nucleosynthesis products
and were
able to rule out a halo origin for this BH binary.
Additionally,
\citealt{fra} claimed that
the surface metallicity of the donor star right before 
the onset of RLO might have been even higher than the observed one,
which would make the argument for a disc origin stronger.
A natal kick seems to be required
for this system and one which exceeds the range of kicks obtained from either
the Hansen \& Phinney or Arzoumanian distributions with a reduction by the ratio
of black hole to neutron star masses.  \\
\indent
From table \ref{tab:min} we see that natal kicks exceeding $70$ km/s are required 
for several systems: 
4U1543-47, 1819.3-2525,
XTEJ1118+480
and 1705-250. These binaries provide us with  evidence that
{\em black holes receive natal kicks of the same size as those received 
by neutron stars}. 
One might have expected the black hole natal kicks to carry the same momentum as
those for neutron stars if, for example, a neutron star formed first (and having received 
a kick) and then a black hole formed later as a result of fall-back material
within the supernova.
In particular, 
the magnitude of the natal kick imparted to the BH
depends on the competition between two timescales:
the fall-back timescale $\tau_{fb}$ and the timescale of the mechanism leading to the natal kick $\tau_{nk}$.
If $\tau_{fb} > \tau_{nk}$,
we expect the fall-back material not to receive the same natal kick has the proto neutron star;
the BH natal kick will then be reduced by the ratio $M_{ns}/M_{bh}$.
In case $\tau_{fb} \lesssim \tau_{nk}$, we expect the BH to receive a full natal kick.
From our simulations it seems that at least in some of the cases
the fall-back material received the same NK as the proto-NS. 
Our result, already strongly suggested in the overall distribution 
of $z$  seen in the binary population synthesis, is surprising.  However, one should note the
large natal kick speeds obtained by the black hole 
in some of the supernova simulations performed
by \citealt{fra}. \\
\indent
We would also like to point out the recent measurement of the distance of the
BH candidate MAXI J1659-152 (\citealt{jon12}). 
Taking the medium value for the distance in the allowed range $d=6\pm 2$ kpc, we derive a distance
from the Galactic plane of  $z\sim 1.7$ kpc. 
This candidate would add to the number of BH binaries found at large distance from the plane 
of the Galaxy (see also \citealt{kul12}), thus likely enlarging the sample of strong candidates 
for BHs receiving large kicks.\\
\indent
A number of mechanisms have been suggested for natal kicks
received by neutron stars, some of which will also apply to black holes,
especially those forming through the subsequent fall-back of material onto 
a proto neutron star. Suggested kick mechanisms include
two main scenarios: hydrodynamically-driven kicks and neutrino-driven kicks.
The former can either be caused by asymmetries in the convective motions under the stalled shock (see \citealt{herant}, \citealt{burro}, \citealt{janka}, 
and the recent work by \citealt{nordhaus})
or by over-stable oscillation modes of the progenitor core (\citealt{gold}). The latter are produced by asymmetries 
in the neutrino flux in a strong magnetic field (\citealt{arras}, \citealt{qian}). Electromagnetically-driven kicks, instead, act once the neutron star has formed:
the off-centre rotating dipole impart the neutron star a kick (\citealt{tademaru}, \citealt{laietal}).
For a review of neutron star natal kicks, see \citealt{lai}.
Alternatively,
if the core is rotating extremely rapidly it may form a 
central object 
surrounded 
by a massive disc on collapse, which may in turn fragment possibly producing
a second compact object orbiting close to the central black hole or neutron
star. This secondary will rapidly spiral-in towards
the primary. If the secondary is a  neutron star, then it may transfer mass
to the primary until it reaches the minimum mass for a neutron star at which
point it will explode potentially giving the primary a kick (\citealt{colpi}, \citealt{mel2}). In case both objects are black holes, then
a merger kick may result from the asymmetric emission of gravitational waves
(see \citealt{ross}, \citealt{zhuge}, \citealt{rasio}, \citealt{ruffert}). In both cases,
the kick received can be several hundreds km/s. 
Or else, when the disc on collapse forms directly around a BH,
it might release its energy as a powerful jet:
if this jets happens to be one-sided,
it might impart a kick to the central BH,
as suggested by
\citealt{barkov}.
The BH-BH merger scenario might be tested
via BH spin measurement. Evidence of highly rotating BHs comes from the
properties of X-ray spectra of Galactic BH binaries (see for example \citealt{zha}, \citealt{laor}
and also \citealt{fen2}).
Ruling out the accretion of matter from a low-mass companion
as origin of the BH spin (see \citealt{mcc}), a BH-BH merger might well fit in this scenario. \citealt{herr}
estimated the recoil velocity for two coalescing BHs of equal mass and opposite and equal spin.
The recoil velocity reaches $\sim 470$ km/s in the extreme case. \citealt{merr} considered the case
of two BHs of unequal mass; they calculated a maximum recoil velocity
of $450$ km/s for a mass ratio $q$ in the range $0.2-0.4$.

\setcounter{table}{4}
\begin{table}
\begin{center}
\caption{$V_{\perp}$ necessary to get to the observed position, and corresponding
minimum natal kick $V_{\rm nk}$ $^a$.}
\label{tab:min}
\begin{tabular}{l c c c c }
\hline
Name& $V_{\perp}$  & $V_{\rm nk}$ & R  & z \\
&\small{(km/s)} & \small{(km/s)} & \small{(kpc)} & \small{(kpc)}\\
\hline3
4U1543-47 &$95$ &$80$  &3.92 &0.70\\
XTEJ1550-564&$22$ &$10$ &5.0&-0.14\\
GROJ1655-40 &$36$ & $0$&4.98&0.13\\
1659-487 &$113$ &-&3.25&-0.67\\
1819.3-2525 &$160$  &$190$ &2.14&-0.82\\
GRS1915+105&$5$ & $0$&6.62&-0.03\\
GS2023+338&$10$ & $0$&7.65&-0.09\\
GROJ0422+32&$25$ & $10$  &9.91&-0.41\\
A0620-003&$10$ & $0$ &8.92&-0.12\\
GRS1009-45&$40$  & $15$ &8.48&0.62\\
XTEJ1118+480& $80$  & $70$ &8.73&1.50\\
1124-683&$50$  & $40$ &7.63&-0.73\\
XTEJ1650-500& $20$ &-&5.71&-0.15\\
1705-250&$420$ & $450$ &0.55&1.36\\
XTEJ1859+226&$80$ &-&7.23&1.20\\
GS2000+251 &$15$ &0&7.21&-0.14\\
\hline 
\end{tabular}
\end{center}
$^a$ For BH-LMXBs that lack strong observational constraints, 
we are unable to calculate accurately $V_{\rm nk}$ so leave it blank here.
\end{table}

\section{Conclusions}
\label{sec:concl}

In this Paper, we have considered the distribution of low-mass X-ray binaries
containing black holes (BH-LMXBs) within the Galaxy as a function of the
distribution of natal kicks given to the black holes.  
We have synthesized a BH-LMXB
population by forming systems randomly throughout the Galactic citealt,
weighted by stellar surface density of the disc. We have given each binary
a mass-loss kick due to the supernova explosion of the primary and added
to this kick a black hole natal kick drawn from one of five kick distributions:
1) a natal kick of zero km/s; 2) one drawn from the Hansen \& Phinney 
distribution (\citealt{hp}); 3) one drawn from the bimodal distribution of
\citealt{arz}; 4) as 2) but with the kick speed
multiplied by the factor $M_{\rm ns}/ M_{\rm bh}$; and 5) as 3) but 
with the kick speed multiplied by the factor $M_{\rm ns}/ M_{\rm bh}$.
We have added the two kicks together with random directions and
combined (randomly) with the original orbital velocity within the Galaxy.
The trajectory of each binary has then been integrated within the Galaxy.

A number of observed BH-LMXBs are found in excess of $1$ kpc from the 
Galactic disc. By comparing our synthesized 
population  to the observed systems, we show that the
hypothesis that black holes only rarely receive a natal
kick is ruled out at very high significance.
The computed distribution is most similar to the observed distribution
when the black hole natal kicks are drawn from the same velocity distribution
as for neutron stars.   
Although we are unable to rule out that black holes receive smaller
kicks than neutron stars, in a number of cases the required natal
kick is very likely to exceed the maximum possible kicks in the 
reduced-velocity distributions (i.e. distributions 4 and 5 above).


\section*{Acknowledgments}

This work was supported by the Swedish Research Council (grants 2008--4089
and 2011--3991). We thank the anonymous referee for helpful comments that improved our paper. SR is thankful to the University of Pavia for the Erasmus grant
and Andrea De Luca for helpful suggestions. MBD and SS gratefully acknowledge the hospitality of the Aspen Center for Physics.

\label{lastpage}
\end{document}